\documentclass[preprint2]{aastex}

\shorttitle{From first stars to the lithium Plateau}
\shortauthors{Piau et al.}

\begin{document}


\title{From First Stars to the Spite Plateau: a Possible Reconciliation
of Halo Stars Observations with Predictions from Big Bang Nucleosynthesis.}


\author{L. Piau \altaffilmark{1}}
\affil{University of Chicago, LASR 933, East 56th street, and JINA: Joint Institute for Nuclear Astrophysics,
Chicago, IL, 60637, USA}
\email{laurent@oddjob.uchicago.edu}

\author{T. C. Beers\altaffilmark{2}}
\affil{Department of Physics \& Astronomy, CSCE: Center for the Study of Cosmic
Evolution, and JINA: Joint Institute for Nuclear Astrophysics,
Michigan State University, East Lansing, MI 48824, USA}
\email{beers@pa.msu.edu}

\author{D.S. Balsara \altaffilmark{3}}
\affil{Department of Physics, University of Notre Dame,
225 Nieuwland Science Hall, Notre Dame, IN, 46556, USA}
\email{dbalsara@nd.edu}

\author{T. Sivarani\altaffilmark{4}}
\affil{Department of Physics \& Astronomy, CSCE: Center for the Study of Cosmic
Evolution, and JINA: Joint Institute for Nuclear Astrophysics,
Michigan State University, East Lansing, MI 48824, USA}
\email{thirupathi@pa.msu.edu}

\author{J. W. Truran \altaffilmark{5}}
\affil{University of Chicago, Department of Astronomy \& Astrophysics, 933, East 56th street, 
Chicago, Il, 60637, USA
Argonne National Laboratory, 9700 South Cass Road, 
Argonne, Il, 60439-4863, USA}
\email{truran@nova.uchicago.edu}
\and
\author{J. W. Ferguson \altaffilmark{6}}
\affil{Physics Department Wichita State University
Wichita, KS 67260-0032, USA }
\email{jason.ferguson@wichita.edu}




\begin{abstract}

Since the pioneering observations of Spite \& Spite in 1982, the constant
lithium abundance of metal-poor ([Fe/H] $ < $-1.3) halo stars near the turnoff has
been attributed to a cosmological origin. Closer analysis, however, revealed that the
observed abundance lies at $\rm \Delta ^7Li \sim$0.4 dex below the predictions
of Big Bang Nucleosynthesis (BBN). The measurements of deuterium abundances
along the lines of sight toward quasars, and the recent data from the Wilkinson
Microwave Anisotropy Probe (WMAP), have independently confirmed this gap. We suggest
here that part of the discrepancy (from 0.2 to 0.3 dex) is explained by a
first generation of stars that efficiently depleted lithium. Assuming that the
models for lithium evolution in halo turnoff stars, as well as the $\rm \Delta
^7Li$ estimates are correct, we infer that between one-third and one-half of
the baryonic matter of the early halo (i.e. $\rm \sim 10^9 M_{\odot}$) was
processed through Population III stars. This new paradigm proposes a very
economical solution to the lingering difficulty of understanding the properties
of the Spite Plateau and its lack of star-to-star scatter down to [Fe/H] $=
-2.5$. It is moreover in agreement both with the absence of lithium in the most
iron-poor turnoff star presently known (HE~1327-2326), and also with new trends
of the Plateau suggesting its low metallicity edge may be reached around
[Fe/H] $= -2.5$. We discuss the role of turbulent mixing associated with enhanced
supernovae explosions in the early interstellar medium in this picture. We
suggest how it may explain the small scatter, and also other recent observational
features of the lithium Plateau. Finally, we show that other chemical properties
of the extremely metal-poor stars (such as carbon enrichment) are also in
agreement with significant Population III processing in the halo, provided
these models include mass-loss and rotationally-induced mixing.

\end{abstract}





\section{Introduction}\label{sec1}

For more than 20 years lithium has been recognized as an efficient probe of the
(early) nucleosynthetic evolution of the Universe. The abundance of $^7$Li in
halo stars is almost independent both of their effective temperatures (between
6400~K and 5600~K) and of their metallicity between [Fe/H] = $ -1.5$ and $-2.5$.
This observational fact was first discovered in the early 1980s by Spite \&
Spite (1982), and has been confirmed by numerous observational studies of
increasing accuracy in the subsequent decades, including : Hobbs \& Duncan
(1987); Rebolo, Beckman, \& Molaro (1988); Spite \& Spite (1993); Thorburn
(1994); Molaro, Primas, \& Bonifacio 1995; Ryan et al. (1996); Bonifacio \&
Molaro (1997); Ryan, Norris, \& Beers (1999); Bonifacio et al. (2002); Asplund
et al. (2006); Bonifacio et al. (2006). The most natural way to understand the
apparently constant $^7$Li abundance on the so-called Spite Plateau is to relate
it directly to production by Big Bang Nucleosynthesis (hereafter BBN). BBN is
understood to produce $^7$Li and the other light species $^2$H, $^3$He and
$^4$He. One clear challenge to this simple picture is the fact that the observed
abundances of $^7$Li in the atmospheres of halo stars are clearly below the
predictions of standard BBN. 

On average, the measurements of $^2$H in intergalactic clouds along the
lines of sight to quasars are $\rm (^2H/H)_p \sim 3 \times 10^{-5}$ (O'Meara et
al. 2001; Burles 2002 and references therein). Strictly speaking, this
estimate is only a lower limit on the primordial $^2$H, as, except for the Big
Bang, there is no astrophysical site currently known to significantly produce
$^2$H. However, the $^2$H observed towards QSOs (in a medium where the
metallicity is below [Fe/H] = $ -1.0$) is probably very close to the actual
primordial value; for various assumptions, the early Galactic disk chemical
evolution models predict no significant change of $^2$H as long as $\rm [O/H] <
-0.5$ (Romano et al. 2006). $3 \times 10^{-5}$ is presumably the
deuterium-to-hydrogen ratio left behind by BBN. In order to have provided such
a deuterium fraction, standard BBN would have to have produced much more lithium
than is currently observed in halo stars on the Spite Plateau. More recently,
the validity of this discrepancy related to lithium found further strong support
through the constraints provided by the cosmic radiation background anisotropies
measured by the Wilkinson Microwave Anisotropy Probe (WMAP). These anisotropies
imply a baryon number $\eta_{\frac{b}{\gamma}}$ around $6 \times 10^{-10}$
(Spergel et al. 2003). In the context of standard BBN, using the most recent
compilation of nuclear reaction rates, this $\eta_{\frac{b}{\gamma}}$ implies a
primordial deuterium fraction in perfect agreement with the $^2$H observations
in intergalactic clouds, but a $^7$Li abundance (herafter referred as A($^7$Li)=
12 + log[N(Li)/N(H)]) much higher than what is observed in halo stars. It is
also interesting to note that these BBN predictions are robust with respect to
constraints on the nuclear cross sections evaluated on the basis of solar
neutrinos (Cyburt et al. 2004), and also with respect to recent estimates of
primordial $^4$He (Olive \& Skillman 2004; Cyburt et al. 2005). The current
situation is therefore the following. On the one hand, the present estimate of
primordial A($^7$Li), based on the most up-to-date nuclear physics, seems
reliably set at A($^7$Li) = 2.6 (Coc et al. 2004). On the other hand, the
observed halo-star lithium plateau appears to be no larger than A($^7$Li) = 2.1
$-$ 2.2, although Mel\'endez \& Ramirez (2004) have argued that changes in the
adopted temperature scale might be able to raise the observed Spite Plateau
value of A($^7$Li) up to 2.4. Fields, Olive, \& Vangioni-Flam (2005) point out
that the revision in the temperature scale suggested by Mel\'endez \& Ramirez
(2004) raises several serious challenges to Galactic cosmic-ray nucleosynthesis
and Galactic chemical evolution, and hence must be considered with caution. 

Lithium is known to be a very fragile element, one that is rapidly destroyed by
nuclear reactions in stellar interiors when the temperature exceeds 2.5 x
$10^6$~K. One possible explanation of the discrepancy between the expected
primordial lithium and that measured on the Spite Plateau could therefore be
that the lithium presently observed in the outer atmospheres of halo stars has
been depleted by stellar-evolution processes over the long history of the
Galaxy. A great number of modeling efforts have investigated this hypothesis (as
recently summarized by Charbonnel \& Primas 2005). As far as we are aware, these
previous works have only addressed the question of whether the halo stars
themselves could be responsible for {\it in situ} lithium depletion during their
main-sequence (hereafter MS) or pre-main-sequence (hereafter pre-MS) evolution.

In this paper we address the question of the early evolution of lithium from a
different perspective, and examine the possible effects of zero-metallicity or
near zero-metallicity (Population III) stars on the abundance of lithium in the
interstellar medium from which (by definition) the next-generation stars formed.
Indeed, no halo star is presently observed that is completely devoid of elements
that were formed post-Big Bang, such as C, N, O, the alpha-elements, and the
iron-peak elements. Hence, all of the recognized halo stars stars must have
formed from material that was enriched -- at least partially-- by a Population
III object or a subsequent stellar generation. If the heavier-than-lithium
element fractions of Spite Plateau stars have been affected, then it stands to
reason that their lithium abundances may have been altered as well, in
particular if the Population III stars are capable of destroying lithium, and
efficiently recycling this Li-depleted gas back into the interstellar medium
(hereafter ISM). In this scenario, other elements, as well as the trends and
observed scatter of other elements, should have been affected too. These
possibilities are considered in turn below.

Implicit in the above scenario for $^7$Li destruction is the assumption that the
$^7$Li-depleted ejecta from massive Population III stars are efficiently mixed
with the proto-Galactic ISM. A static ISM would not permit efficient mixing.
However, several lines of evidence, cataloged in later sections, suggest that
the proto-Galactic ISM was in fact quite dynamic. Recent observations of distant
ultraluminous infrared galaxies (ULIRGs) (Shapley et al. 2001; Daddi et al.
2005; Yan et al. 2005) show that a significant amount of star formation might have
taken place as early as $z \sim 6$. Numerical simulations of $\Lambda$-dominated
cold dark matter cosmologies (Nagamine et al. 2004; Night et al. 2005) also
support scenarios where galaxies form early. Furthermore, they support scenarios
where larger systems formed via the coalescence and mergers of smaller systems,
a scenario that also finds some observational support (Dasyra et al. 2006).
Thus, several lines of evidence indicate that proto-galactic ISMs were
dynamically evolving due to massive Population III star formation or mergers in
the early Universe. Perhaps the best evidence to indicate that our Galaxy might
also have passed through such a phase of evolution emerges from the small
scatter in alpha-capture elements that has been observed in very metal-poor halo
dwarfs down to [Fe/H]$\sim -2$ (Norris et al. 2001, and references therein).
The fact that a wide range of $\alpha$-capture elements have evolved in
lock-step with [Fe/H], and presently exhibit extremely small observed scatter
about well-defined trends with [Fe/H], suggests that massive stars not only
produced these elements, but also that the winds and Type II SNe from these
stars were very efficient at dispersing metals throughout the proto-Galactic
ISM. Below [Fe/H]$\sim -2$ the halo stars still exhibit a small scatter in
$\alpha$-capture or iron-peak elements (Cayrel et al. 2004; Arnone et al. 2005).
However, the r-process elements seem to be more scattered (Truran et al. 2002),
which makes an efficient mixing of the ISM less probable in these metallicity
regimes. Studies of supernova-driven mixing of the ISM have already been
presented in the literature (Korpi et al. 1999; Avillez \& Breitschwerdt 2004;
Balsara et al. 2004; Balsara \& Kim 2005; Mac Low et al. 2005). We draw on the
insights available from the literature to understand the dispersal of
lithium-depleted but metal-enriched material throughout an early ISM.

\begin{figure}[Ht]
\begin{center}
\includegraphics[angle=0,scale=.35]{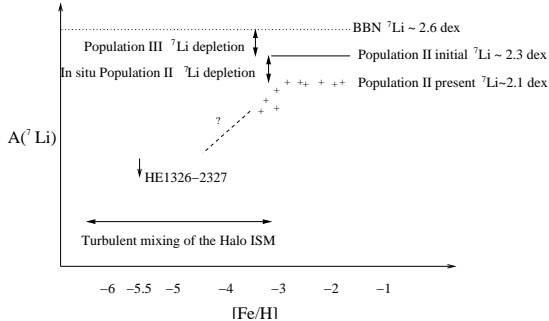}
\caption{A schematic of the $^7$Li vs. [Fe/H] relationship and the evolution of the $^7$Li abundance we propose
occured in the early halo. The dotted-line shows the BBN $^7$Li abundance. The solid line shows the
initial Population II $^7$Li abundance. The crosses show the abundance in $^7$Li 
presently observed in the halo. The dashed-line is indicative of a possible
decrease in the $^7$Li abundance from the apparent edge of the Spite Plateau 
around [Fe/H]$\sim -2.5$ to objects similar to HE~1327-2326.}
\label{fig1} 
\end{center}
\end{figure}

For discussion of the lithium discrepancy, we focus on the metal-poor stars with
effective temperatures above 6000 K. These objects are less likely to have
undergone significant lithium depletion themselves, at least as compared to
their cooler counterparts, due to the smaller outer convection region with
increasing mass (and hence temperature). In \S \ref{sec2} we address the
question of whether or not these main-sequence-turnoff stars could have depleted
a substantial fraction of their initial lithium content when non-standard mixing
processes are considered (in addition to convection, microscopic diffusion
and gravitational settling). We find that these stars probably experienced a
moderate lithium depletion of their surface abundances ($\sim$ 0.2 to 0.4 dex ). Section
\ref{sec3} discusses the possible role that Population III stars might have
played. We estimate that between one-third and one-half of the Galactic halo
matter must have been processed through these stars in order to explain the
identified discrepancy between the lithium plateau observations and the
predicted level of primordial BBN lithium production. We emphasize that this
derived fraction is a result of our modeling, and not an input assumption. To
arrive at this conclusion, we make the assumptions: (a) A($^7$Li) from BBN is
2.6, (b) the level of observed Spite Plateau lithium is between 2.1 and 2.2
for main-sequence-turnoff stars with metallicity above [Fe/H] = $-2.5$, and (c) that
present models of Population II stars correctly predict the likely amount of
{\it in situ} lithium depletion in these stars. In \S \ref{sec4} we discuss interesting
new lithium trends that have been observed for halo stars below [Fe/H]= $ -2.5$.
In \S \ref{sec5} we present arguments in support of the likelihood of efficient
early mixing of the ISM due to turbulent mixing by early supernovae. Figure
\ref{fig1} summarizes the general features of our model describing lithium
history, from the BBN to present observations in the halo. Note that this Figure
indicates the presence of two distinct contributions to primordial lithium
depletion in order to reach the presently observed Spite Plateau value of
lithium abundance.

In \S \ref{sec6} we summarize other recent observations that might be
explained by the scenario we propose, including (a) the unique abundance patterns observed
among stars of the lowest metallicity, (b) the production of primary nitrogen
and the observed trends of C/O and N/O in very low-metallicity stars, (c) the
lack of observed star-to-star scatter in alpha- and iron-peak elements in stars
of very low metallicity, (d) the peculiar features of carbon in extremely-metal-poor stars
and hyper metal-poor stars\footnote{Following the taxonomy suggested by Beers \&
Christlieb (2005), the extremely metal-poor star term refers to stars with
[Fe/H]$ < -3$, while the hyper metal-poor star term refers to the stars with
[Fe/H]$ < -5$.}. Our conclusions and discussion follow in \S \ref{sec7}, where we
also suggest several testable predictions of the Population III processing
model. The reader who is only interested in the main idea of this article might
read \S 3 and \S 4, and then directly move to the conclusions presented in
\S \ref{sec7}.

\section{Lithium Evolution in Very Metal-Poor Stars}\label{sec2}

\begin{figure}[Ht]
\begin{center}
\includegraphics[angle=90,scale=.35]{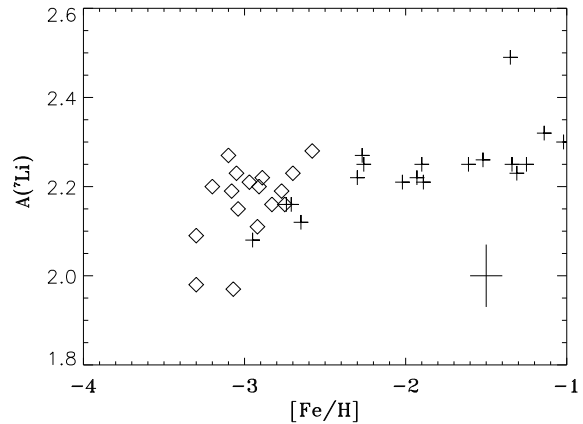}
\caption{A($^7$Li) measurements among the very low-metallicity halo turnoff stars. Crosses:
data from Asplund et al. (2006), Diamonds: data from Bonifacio et al. (2006).
The abundances from the two samples assume the same $\rm T_{eff}$ scale, so
they can be compared directly. The error bars shown on the lower right 
represent the (comparable) uncertainties
on [Fe/H] and A($^7$Li) from both studies. Three very low-metallicity
stars from Bonifacio et al. (2006) clearly appear below the plateau.}
\label{fig2} 
\end{center}
\end{figure}

Several observed facts about the nature of the Spite Plateau are of central
importance for testing our Li-astration model. The lithium plateau is flat only
as a first approximation. Ryan et al. (1996) have argued for the presence of
small, but significant, correlations in A($^7$Li) with both effective
temperature (+0.04 dex per 100~K) and with metallicity (+0.11 dex per dex). In
an effort to avoid any correlations with temperature, Ryan et al. (1999)
investigated a sample of 23 halo dwarfs with similar $\rm T_{eff}$ (between
6100~K and 6300~K), and obtained a slope of A($^7$Li) with respect to [Fe/H] of
+0.12 dex per dex, quite similar to the previous analysis. This slope was
suggested by Ryan et al. (2000) to be the signature of lithium production by
spallation (see Fields \& Olive 1999) from cosmic rays and the supernova
$\nu$-process (Woosley et al. 1990; Woosley \& Weaver 1995), which gradually
increase the lithium content of the interstellar medium from which successive
generations of stars formed. Recent observational results suggest that the slope
of A($^7$Li) with [Fe/H] sharply changes below [Fe/H] $\sim -2.5$ (Asplund et
al. 2006). In this very low-metallicity regime the increase of the lithium
fraction with metallicity seems more rapid, as shown by Figure \ref{fig2}. It is
also possible that the scatter of A($^7$Li) similarly goes up below this
metallicity (Bonifacio et al. 2005, 2006). The star-to-star scatter in measured
A($^7$Li) for metal-poor stars on the plateau is {\it extremely} small, on the
order of 0.03-0.05 dex (Ryan et al. 1999; Asplund et al. 2006; Bonifacio et al.
2006), well within the expected observational errors\footnote{We note that
debate continues on whether or not the Spite Plateau exhibits a slope (on the
order of 0.1-0.2 dex per dex) in A($^7$Li) vs. [Fe/H], with several authors
coming out in favor of such a slope (Ryan et al. 1999; Asplund et al. 2006), and
others not (Mel\'endez \& Ramirez 2004; Bonifacio et al. 2006).} However, Figure
\ref{fig2} suggests a possible increase in the scatter of A($^7$Li) below [Fe/H]
$\sim -2.5$. Clearly, these new features concerning the break in the the slope
and increased scatter of the Spite Plateau for very low-metallicity stars
require verification based on measurements of lithium abundances for additional
larger samples of stars with extremely low metallicity. We remark that the
tendency for the lithium fraction to decrease strongly in (at least some) dwarfs
below [Fe/H] $\sim -2.5$ is confirmed by high-resolution observations of the the
hyper metal-poor star HE~1327-2326, where an upper limit for A($^7$Li) of 1.5
dex is obtained (Frebel et al. 2005; Aoki et al. 2006a). Very recent
results, yet to be published, further show that $\rm A(^7Li)<0.9$ should be
adopted in HE~1327-2326 (A. Frebel, private communication).

We now consider the evolution of lithium abundances for extremely metal-poor
turnoff stars with [Fe/H] = $-3.5$ having the typical distribution of metals
associated with most halo stars (hereafter composition A): [$\alpha$/Fe]= +0.35,
where $\alpha$ stands for oxygen and the $\alpha$ elements (Ne, Mg, Si, S, Ar,
Ca, and Ti), while [C/Fe]= [N/Fe] = 0 (see Piau 2006 for details). Similarly, we
consider the lithium abundance evolution in stars with the same composition as
HE~1327-2326 (hereafter composition B) which, with [Fe/H]= $-5.5$ is the most
iron-poor star presently known (Frebel et al. 2005). HE~1327-2326 has a rather
unique composition, and in particular exhibits huge carbon, nitrogen, and oxygen
content with respect to its iron content. Depending on the (still unclear)
evolutionary status of this star, [O/Fe] lies between +2.8 $\pm 0.2$ (subgiant)
and +2.5 $\pm 0.2$ (dwarf), while [C/Fe] = +4.1 (Frebel et al. 2006). Both of the
compositions we model are assumed to have the primordial helium mass fraction
$\rm Y_p=0.2479$, as well as primordial $^7$Li and $^2$H number fractions of
$4.15 \times 10^{-5}$ (A($^7$Li) = 2.6) and $2.60 \times 10^{-5}$ respectively, as estimated
by Coc et al. (2004). The modeling of the evolution of $^7$Li abundances is
performed herein using the stellar evolution code CESAM (Morel 1997). The
general inputs to the code (equation of state, opacities, convection modelling,
etc.) are similar to the description provided in Piau (2006), and we do not
repeat them here. 

\begin{table*}[Ht]
  \begin{center}
    \caption{Initial Metal Fractions Relative to [Fe/H] Adopted in our Models} 
\vspace{1em}
    \renewcommand{\arraystretch}{1.2}
    \begin{tabular}[h]{lcc}
      	Metal         & ``Typical'' Halo Star  &    HE~1327-2326  \\
      \hline	                                             
      	              &   Composition A      &   Composition B    \\
      \hline	                                             
	[C/Fe]        &   0                  &   +3.9             \\
      \hline	                                             
	[N/Fe]        &   0                  &   +4.2             \\
      \hline	                                             
	[O/Fe]        &   0.35               &   +2.6             \\
      \hline	                                             
	[Fe/H]        &   $-$3.5             &   $-$5.3           \\
      \hline  
	Y, Z          &   0.2479,  $\rm 9.77 \times 10^{-6}$  &   0.2479, $\rm 2.98 \times 10^{-4}$    \\
      \hline
\tablecomments{We adopt the Grevesse \& Noels (1993) solar composition to infer
absolute abundances. Note that the accuracy on heavy element abundances for
HE~1327-2326 are on the order, or larger, than the change of abundances
resulting from the revision of the solar composition by Asplund et al. (2006). Y
and Z stand, as usual, for the total mass fractions in helium and the heavy
elements.} 
\end{tabular}
\label{tab1}
  \end{center}
\end{table*}

\begin{figure}[Ht]
\begin{center}
\includegraphics[angle=90,scale=.35]{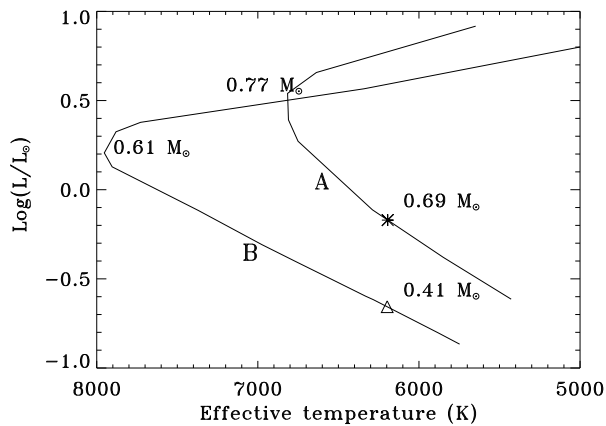}
\caption{Comparative isochrones at 13.5 Gyr for compositions A and B models without diffusion. 
The positions and masses of the objects exhibiting the present HE~1327-2326 $\rm T_{eff}$ are
indicated by a triangle and a star on each isochrone. The turnoff masses on each
isochrone are also indicated.}
\label{fig3} 
\end{center}
\end{figure}

Table \ref{tab1} summarizes the initial compositions we have adopted in our
models. In the case of HE~1327-2326 all of the initial metal abundances were
increased by $\sim$ 0.2 dex in order to roughly take into account the
microscopic diffusion and gravitational settling of heavy elements
(therefore [Fe/H] = $-5.3$ initially) during the main-sequence evolution
(herafter MS) of halo dwarfs. Because of its very thin outer convection
zone, it is probable that HE~1327-2326 experienced a stronger diffusion and
settling of the heavy elements than a typical halo star. Thus, we also
constructed a model of HE~1327-2326 where the initial [Fe/H] is increased to
$-4.5$ (see below). Because of the small impact of metals on the Equation of
State (EOS) in the Population II regime, the main difference in the modeling
between the typical halo-stars' composition and that of HE~1327-2326 relies on
the opacities. The required opacities have been taken from the OPAL web site
(http://www-phys.llnl.gov/Research/OPAL/) for halo-star composition (the
$\alpha$-element-enhanced table of F. Allard) or alternatively, generated online
on this site for HE~1327-2326. The low-temperature opacities ($\rm log\; T <
3.75$) were generated by one of the authors (see Ferguson et al. 2005). In order
to achieve $\rm T_{eff}=6200$ K at an age of 13.5 Gyr, we predict the mass of
HE~1327-2326 to be 0.41 $\rm M_{\odot}$. This is much smaller than the mass of a
similar-temperature and similar-age star with the typical composition of an
extremely metal-poor halo star having [Fe/H] = $-3.5$: 0.69 $\rm M_{\odot}$. It
is well known that the effective temperature strongly increases, at a given mass
and age, when the metallicity decreases. More precisely, this effect is
related to the drop of abundance of the metals having low-energy
first-ionization potentials, such as Na, Ca, and Fe (e.g., Gehren et al. 2004).
The less numerous these metals, the less abundant the H$^-$ ion, and the smaller
the atmospheric opacity. In the case of HE~1327-2326, this trend is reinforced
by the presence of a convective core associated with $^{12}$C to $^{14}$N
conversion during most of the MS phase. The combination of the low luminosity
and the high initial carbon fraction allows this convective core (induced by
CN-cycle burning) to survive to the age of 15 Gyr, which is when we stop our
computations. Due to the convective core in models with HE~1327-2326
composition, they live longer on the MS than typical halo stars. The impact of a
C-burning convection zone on the MS lifetime has been mentioned in previous
stellar evolution studies (e.g., Maeder \& Meynet 1989). More generally, the
increasing size of the central regions lengthens the MS duration, because they
increase the amount of available hydrogen fuel (see, e.g., Schaller et al.
1992). Table \ref{tab2} shows the main properties of our $\rm T_{eff} \sim 6200$
K models at 13.5 Gyr. Because stars with similar composition to HE~1327-2326 may
be detected in the near future, Figure \ref{fig3} shows comparative isochrones
of this peculiar-composition star as compared to typical extremely metal-poor
halo stars.

\begin{table*}[Ht]
  \begin{center}
    \caption{Initial Mass, Effective Temperature, Temperature at the Base of the outer Convection Zone, 
Mass of Outer Convection Zone, Bolometric Luminosity, Surface Gravity, 
and Central Hydrogen Mass Fraction ($\rm X_c$) for Objects of the
Two Compositions Considered}
\vspace{1em}
    \renewcommand{\arraystretch}{1.2}
    \begin{tabular}[h]{lccccccc}
      	Composition   &  Mass ($\rm M_{\odot}$) &  $\rm T_{eff}$ (K) &  $\rm
T_{BCZ}$ (K) &  $M_{CZ}$ ($\rm M_{\odot}$)    & L($\rm L_{\odot}$) & log g  &  $\rm X_c$  \\
      \hline				         		                  
	A             &  0.69                   &    6194            &   $1.2 \times 10^6$     &  $4.5 \times 10^{-3}$      & 0.675              & 4.57   &  0.14   \\
      \hline				         		                  
	B             &  0.41                   &    6197            &   $4.8 \times 10^5$     &  $2.9 \times 10^{-4}$      & 0.220              & 4.83   &  0.54   \\
      \hline				         
\tablecomments{The presence of a convective core translates into a much higher
$\rm X_c$ in the case of composition B (HE~1327-2326). We do not take into
account the possiblity of core-convection overshooting; this mechanism would
increase $\rm X_c$ even further.} 
\end{tabular}
\label{tab2}
\end{center}
\end{table*}

Lithium is a fragile element, and therefore it can be destroyed even in the
envelopes of low-mass stars. It is difficult to accurately model its history,
mainly because it probably depends on non-convective mixing mechanisms, as
suggested by the lithium evolution of solar-like stars. The observational facts
are: (1) lithium has been depleted in the Sun by a factor $\sim$ 160
(Asplund et al. 2005), and (2) there appears to be depletion in solar analogs on
a timescale of a few hundred Myr, as shown by studies of Galactic open clusters
(Sestito \& Randich 2005). This depletion occurs in objects having, as does our
Sun, the base of their convection zones (hereafter BCZ) at temperatures below
2.5 10$^6$ K, and therefore is related to non-standard mixing processes in the
sense that they are not the signature of convection only, but must involve
rotation effects, magnetic effects, or internal wave-induced mixing. Our
computations include two possible non-standard mixing processes. The first is
the so-called tachocline mixing process (Spiegel \& Zahn 1992). This
rotationally-induced mixing occurs in a thin layer below the convection zone. It
correctly predicts the timescale for $^7$Li depletion as seen in Galactic open
clusters (Brun et al. 1999; Piau et al. 2003) as well as the amount of $^7$Li
depletion in the Sun. Moreover, this process preserves the initial solar $^9$Be,
and provides better agreement between the theoretical solar sound-speed profile
and that inferred from helioseismology (Brun et al. 1999). There are two
important parameters in the tachocline modeling: (1) the assumed width of the
tachocline region, which we take to be $2.5 \%$ of the stellar radius, based on
solar seismic measurements (Corbard et al. 1999, Charbonneau et al. 1999), and
(2) the buoyancy frequency assumed in the tachocline region, which we assume to
be 10 $\mu$ Hz (for discussion about these choices we refer the reader to Piau
et al. 2003). Next, one needs to introduce the rotation history into the
modeling. Since this history is unknown in Population II stars, we assume it is
similar to the Population I low-mass objects. We consider that the star rotates
as a solid at all times, and that the initial rotation period is 8 days. Despite
the contraction we assume the angular velocity remains unchanged during the
first million years of evolution, because the star is coupled to its early
circumstellar disk. Past this age the star spins up. The angular momentum is
roughly constant, but the object continues to contract. After the Zero-Age
Main-Sequence (ZAMS) stage, it slows
down through the effect of magnetic wind braking. We follow the prescriptions
given by Kawaler (1988) on the losses of angular momentum, J :

\begin{equation}
\rm \frac{dJ}{dt}=-K{\Omega}^3\,\rm ,if  \, \Omega < {\Omega}_{sat} \,\, and \,\, \frac{dJ}{dt}=-K{\Omega}{{\Omega}_{sat}}^2\,\rm , if \, \Omega > {\Omega}_{sat}
\label{eq1}
\end{equation}

\noindent In the above, $\rm \Omega$ is the angular velocity, $\rm \Omega_{sat}$ is 
a saturation threshold in angular velocity, and K is a constant. The parameter $\rm
\Omega_{sat}$ is set to $14 \Omega_{\odot}$, following the observations of the
rotation history of solar analogs in open clusters (Bouvier, Forestini, \&
Allain 1997). The constant K is either adjusted to $2.0 \times 10^{46} \rm g\,s\,cm^2$
(composition A and 0.69 $\rm M_{\odot}$ star) or $8.5 \times 10^{45} \rm g\,s\,cm^2$
(composition B and 0.41 $\rm M_{\odot}$ star) in order to let our models rotate
at 30 $\rm km/s$ at the Hyades age ($\sim$ 625 Myr). This is the velocity
assessed by Gaig\'e (1993) from observations of objects having $\rm T_{eff} \sim
6200$ K in this cluster. The impact of the combined microscopic diffusion 
gravitational settling and tachocline mixing have been explored in detail by
Piau (2006) for halo stars of metallicity [Fe/H]= $-2.0$. These models indicate
that lithium depletion at the MS turnoff for the Spite Plateau is $\sim$ 0.2
dex. In the case of $^7$Li, this depletion always occurs because of
microscopic diffusion, not through nuclear destruction. However, in the case of
$^6$Li and $\rm 0.69 M_{\odot}$ models of composition A, some nuclear
destruction occurs.

The other possibility we explore for non-standard mixing is the prescription
adopted by Richard, Michaud, \& Richer (2005). Following these authors, we have added
turbulent mixing below the convection zone with a turbulent diffusion
coefficient $\rm D_t$ :

\begin{equation}
\rm D_t=400 D_{He}(T_o)\frac{\rho(T_o)}{\rho}^3
\label{eq2}
\end{equation}

\noindent where $\rm D_{He} (T_o)$ and $\rm \rho (T_o)$ are, respectively, the helium microscopic diffusion 
coefficient and density in a star at $\rm T_o = 10^6$ K. This turbulent mixing
law is the most satisfactory one among the various prescriptions for turbulent
mixing explored by Richard et al. (2005), in the sense that it produces the best
agreement to the features of the lithium Spite Plateau without inducing $^6$Li
depletion. This point is now supported by the recent observations of $^6$Li on
the Spite Plateau (Asplund et al. 2006). Similar to the models of Piau (2006),
the results from Richard et al. (2005) suggest that the lithium depletion is
around 0.2 dex near the MS turnoff for the turbulent mixing law of equation
\ref{eq2}.

Table \ref{tab3} lists our results for both the extremely metal-poor 
Population II star (composition A) and for HE~1327-2326 (composition B). 
The masses of the stars ending up with the 
required effective temperature at 13.5 Gyr slightly depend on the diffusion 
processes that are taken into account. For instance, a model of a given mass ends up with a 
smaller effective temperature when the microscopic diffusion and gravitational settling are taken into account
than when they are not. For this reason the models in Table 
\ref{tab3} correspond to a small range of masses. In both composition A and B
models no pre-MS $^6$Li or $^7$Li depletion is expected if we consider
the objects now at 13.5 Gyr and having $\rm \sim T_{eff}=6200 K$.
Subsequently, a MS depletion occurs, on the order of 0.06 dex to 
0.8 dex for $^7$Li. Similar to what happens on the
Spite Plateau stars near the turnoff, this is a diffusion and settling
effect, and not the result of nuclear destruction from deep mixing because, 
as illustrated in Table \ref{tab2}, the 
temperature near the BCZ is well below the $^7$Li destruction temperature. The BCZ temperature
is moreover unaffected by the turbulent mixing prescription below the convection zone. 
A few additional remarks need to be made:

For composition A: $^6$Li evolves during MS because of nuclear destruction, in
contrast to $^7$Li. For these models the base of the outer convection zone lies
close enough to the regions where the temperature exceeds $2 \times 10^6$~K, and
material is dragged down to these regions by non-standard mixing. Indeed,
$T_{BCZ}=1.5 \times 10^6$~K at 8 Gyr, and this temperature decreases slowly towards
the end of the MS to the value reported in Table \ref{tab2}. The depletion
therefore is stronger for $^6$Li than it is for $^7$Li.

For composition B: the depletion is slighly less important for $^6$Li than for
$^7$Li, as $^7$Li is heavier, and the depletion results not only from
microscopic diffusion but also from gravitational settling for both isotopes.
Moreover, the tachocline mixing models that do not include the microscopic
diffusion and gravitational settling show $\sim$ 0.6 dex more $^7$Li than those
that include these effects. This is a hint that the microscopic diffusion and
gravitational settling and their interplay with the turbulence must be carefully
taken into consideration in future studies of hyper metal-poor stars. As the
evolution of $^7$Li suggests that the heavy-element surface abundances could
have dropped significantly during the lifetime of HE~1327-2326, we also computed
a model of this star starting with $\rm [Fe/H] = -4.5$ and $\rm [X/Fe]$ values
otherwise identical to table \ref{tab1}. This model includes the diffusion and
tachocline mixing effects. Its $^7$Li and $^6$Li chemical abundances at 13.5 Gyr
and are very similar to what is reported in table \ref{tab3} with $\rm A(^7Li)
=1.81$ and $\rm A(^6Li)=0.42$.

HE~1327-2326 is not the first metal-poor halo turnoff star that is also lithium
poor. Indeed, similar objects have been known for some time (Ryan et al. 2002).
They represent between 5 \% and 10 \% of the Spite Plateau stars.
However, these objects have iron abundances that are much higher than
HE~1327-2326, and rotate faster than typical Spite Plateau stars (Ryan \& Elliott
2005). Moreover, they generally belong to binary systems: three out of four of
these Li-poor stars in the Ryan et al. (2002) sample are confirmed binaries, with
the secondary suspected to be a compact object. A fourth member of the Ryan
et al. Li-poor sample of stars, CD-31:19466, however, does not presently appear
as a binary or as a fast rotator. In this respect, it is comparable to
HE~1327-2326 because of it low lithium abundance, despite having a much higher iron
content. In the other Li-poor objects, binarity and rotation suggest that the
surface composition and angular momentum might have been affected by accretion
from the companion (see Piau 2006 for a discussion). This is in contrast to
HE~1327-2326 where we note that the rotation is presently estimated to be
similar to the lithium plateau stars and no clues of binarity have been detected
so far (Aoki et al. 2006a; Frebel et al. 2006).

\begin{table*}[Ht]
  \begin{center}
    \caption{Initial Lithium Isotope Abundances From Turbulent Mixing Models} 
\vspace{1em}
    \renewcommand{\arraystretch}{1.2}
    \begin{tabular}[h]{lccccc}
      	Composition     & Microscopic Effects   &   No Microscopic Effects   &  Microscopic Effects   \\
      	and Mass Range  & \& Tachocline Diffusion &    and Tachocline Diffusion   &   and ``Richard'' Diffusion  \\
      \hline				         
	A, 0.69-71 $\rm M_{\odot}$ & $^7$Li=2.38, $^6$Li = $-0.01$ & $^7$Li=2.54, $^6$Li=0.06  &    $^7$Li=2.26, $^6$Li=0.00 \\
      \hline			         
	B, 0.41-43 $\rm M_{\odot}$ & $^7$Li=1.83, $^6$Li=0.44 &  $^7$Li=2.52, $^6$Li=1.02  &    $^7$Li=1.96, $^6$Li=0.54 \\
      \hline				         
\tablecomments{The masses of the models achieving $\rm T_{eff}=6200$~K at 13.5 Gyr vary
slightly as a function of the diffusion processes considered. The models
including the microscopic, gravitational settling, and tachocline diffusion are
the heaviest ones, with 0.71 and 0.43 $\rm M_{\odot}$ for composition A and B,
respectively. The models without microscopic diffusion and gravitational
settling are the lightest ones, with 0.69 and 0.41 $\rm M_{\odot}$ for
composition A and B, respectively. Microscopic effects include microscopic
diffusion and gravitational settling. The ``Richard'' diffusion refers to the
turbulent diffusion explained in the text. All of the models were started with
A($^7$Li) = 2.6 and A($^6$Li) = 1.1 }
\end{tabular}
\label{tab3} 
\end{center}
\end{table*}

In conclusion, it is very unlikely that HE~1327-2326, as well as a $\rm
T_{eff}=6200 K$ and composition A object have decreased by more than an
order of magnitude their initial surface lithium in the course of their
evolution. HE~1327-2326 most probably exhibits its pristine abundance to within
0.8 dex. For a composition A object at $\rm T_{eff}=6200$~K, the $^7$Li
depletion we compute is 0.4 dex at most. This conclusion is unchanged for
reasonable age or effective temperature variations. The results displayed in
Table \ref{tab3} are rather insentive to age uncertainties: a change from 13.5
to 11.5 Gyr in the age of the models would increase A($^6$Li) or A($^7$Li) by
less than 0.2 dex in any of the cases considered. Similarly, a slightly incorrect
measurement of $\rm T_{eff}$ would not change the global picture -- an error of
100~K in $\rm T_{eff}$ translates to an error of $\sim$ 0.06 dex on A($^7$Li) for
the considered metallicity and effective temperature range (P. Bonifacio, private
communication). Models with effective temperatures slightly above or below 6200~K
would not change the results either. For instance, in the case of models
including microscopic diffusion, gravitational settling, and tachocline
diffusion, but with $\rm T_{eff}$=6000~K, A($^7$Li)= 2.22 and 1.82 at 13.5 Gyr for
composition A and B, respectively. The variations of A($^7$Li) with $\rm T_{eff}$
are smooth around the turnoff. Interestingly, in the study of Bonifacio et al.
(2006), the star CS~22948-093 ($\rm T_{eff}=6356$~K, [Fe/H]= $ -3.30$) appears
to have A($^7$Li) = 1.98, slightly below the lithium plateau, estimated to be
around A($^7$Li) = 2.17 by these authors. This star could be compared to our
composition A and $\rm 0.69 M_{\odot}$ models. Its observed $^7$Li abundance is
at least 0.3 dex below our predictions, which also suggests that this extremely
metal-poor star inherited its initial lower lithium fraction from its progenitor
cloud. 

\section{Lithium Depletion and Astration in Population III stars}\label{sec3}

In the model we are exploring, the presently observed lithium abundance for
metal-poor stars on the Spite Plateau represents (or nearly represents) the
composition of the ISM out of which these low-mass stars originally formed (as
shown in \S 2). If we posit that the original composition of the ISM from which
high-mass zero-metallicity stars formed is that given by predictions of BBN,
then we must look to such progenitor stars as the source of not only the initial
iron and other elements among subsequently formed low-mass stars, but we must
also consider the consequences on the lithium abundance of the ISM after these
stars have ejected their processed material. 

\begin{figure}[Ht]
\begin{center}
\includegraphics[angle=90,scale=.35]{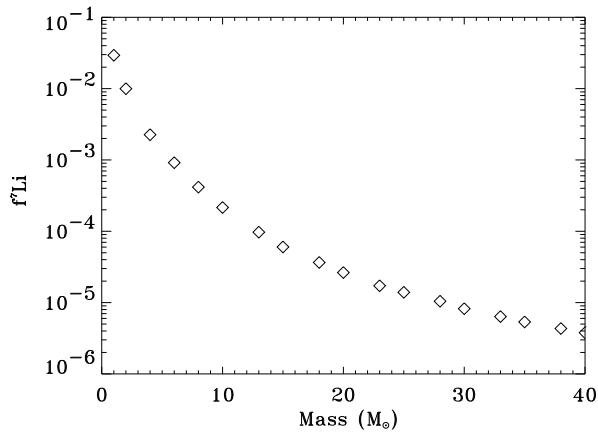}
\caption{Fraction of the mass, $\rm f_{^7Li}$, in the outer layers, where $^7$Li is not depleted in Population III
objects, as a function of the total mass.}
\label{fig4} 
\end{center}
\end{figure}

Using the CESAM code, we have constructed standard models of stars totally
devoid of metals in the $1$ to $\rm 40\;M_{\odot}$ mass range. The hydrogen mass
fraction is set to 0.75 and, correspondingly, the helium fraction is set to
0.25. Neither rotation nor diffusion effects are considered. The models are
evolved until the central temperature reaches $10^8$~K, or the age is 100 Myr,
whichever comes first\footnote{All of our models above $\rm 6 M_{\odot}$ are
still contracting when they reach a central temperature of $10^8$ K. Due to the
absence of CNO, and the lower efficiency of proton-proton chains in these
temperature regimes, these Population III models contract until they ignite the
triple-$\alpha$ reaction, which subsequently allows the CNO cycle to begin (Ezer
\& Cameron 1971).}. We are not interested here in the evolutionary effects
associated with these Population III objects, but on their ability to
deplete their initial lithium content. The outer layers of the Population III
stars we model are radiatively stable, or for lower-mass models, they exhibit
very shallow outer convective regions. For example, the $\rm 1 M_{\odot}$ model
exhibits convective instability over only the outer $0.75\%$ of its radius.
Thus, the material present in the outer regions of such stars will, {\it a
priori}, not be exposed to temperatures above the lithium depletion temperature of $\rm
2.5 \times 10^6$~K. Figure \ref{fig4} shows the fractional mass, $\rm f_{^7Li}$, where
lithium is preserved as a function of the total mass of the star. To evaluate
these stellar mass fractions, we simply considered regions of the star with
temperature below $\rm 2.5 \times 10^6$~K. As in finite-metallicity stars, Population
III stars clearly destroy most of their initial $^7$Li. This trend increases
with increasing mass.

The global impact on $^7$Li in the generation of stars that directly follow the
very first generation will depend on the initial mass function (hereafter IMF)
of these first stars. Following Ballero et al. (2005), we assume the IMF of
Population III stars is: $$\rm \phi(m) \propto m^{-2.7},\, for \, M_l \leq M
\leq M_u .$$ This IMF, together with our $\rm f_{^7Li}$ verses mass relation,
leads us to an estimate of the average mass, $\rm \bar{m}_{^7Li}$, where $^7$Li is
undepleted for every solar mass of matter processed through Population III
stars:

\begin{equation}
\rm \bar{m}_{^7Li}=\frac{\int_{M_l}^{M_u} f_{^7Li}m^{-1.7}dm}{\int_{M_l}^{M_u} m^{-1.7}dm}
\label{eq3}
\end{equation}

Considering a lower- and upper-mass range of $\rm M_L = 1\; M_{\odot}$ and $\rm
M_U = 40\;M_{\odot}$, respectively, equation \ref{eq3} leads to an average $\rm
\bar{m}_{^7Li}=6.9 \times 10^{-3}\; M_{\odot}$. We believe that our adopted upper mass
cut is well justified. Very massive stars ($\rm M_{\star} > 100\;M_{\odot}$) in
the first generation have been suggested (Oh et al. 2001). The presence of such
massive stars is not, however, supported by the observed compositions of very
and extremely metal-poor stars (lack of an observed odd-even pattern), nor are
they required for reionization of the intergalactic medium (Tumlinson,
Venkatesan, \& Shull 2004) HE~1327-2326 and HE~0107-5240 ([Fe/H] = $-5.4$;
Christlieb et al. 2002) are both carbon-rich stars, while HE~1327-2326 is also
barium rich. Recent models of very massive stars, however, yield no significant
carbon or r-process elements (Heger \& Woosley 2002). Contrary to our $\rm M_u$,
which we believe is plausible, a higher $\rm M_l$ around $\rm 10\;M_{\odot}$ is
more likely, because of the absence of efficient coolants of the ISM, such as C
and O at very low or zero metallicity (Bromm \& Loeb 2003). If we take $\rm M_l
= 10\;M_{\odot}$ and $\rm M_u = 40\;M_{\odot}$, we evaluate $\rm \bar{m}_{^7Li}=
6.7\, 10^{-5} M_{\odot}$ from equation \ref{eq3}. In either case ($\rm M_L = 1\;
M_{\odot}$ or $\rm M_L = 10\;M_{\odot}$), only a very small fraction of the
astrated material is not depleted in lithium.

How does $\rm \bar{m}_{^7Li}$ relate to the amount of $^7$Li left behind by the
first generation of stars? One could argue that stellar evolution is able to
produce $^7$Li, so that our $\rm \bar{m}_{^7Li}$ could be a lower limit.
However, it is presently difficult for suggested early lithium production
processes to produce an amount of $^7$Li at a level of $\sim$ 2. Out of the
three processes that are usually invoked for lithium production by stars, two
are related to intermediate- or low-mass stars, the hot bottom burning process
in red/Asymptotic Giant-Branch (AGB) stars (Sackmann \& Boothroyd 1992; Sackmann
\& Boothroyd 1999), and novae explosions (Hernanz et al. 1996). The
low-mass objects are responsible for the transition in the typical chemical
compositions from halo to disk stars. This occurs in a metallicity regime much
higher than the one we are dealing with here. The (massive) AGB stars could 
affect Galactic chemical history earlier, 
because they have evolutionary timescales as short as 40 Myr.
But the present Galactic chemical evolution models suggest that AGB stars have a
negligible impact on $^7$Li history before a metallicity [Fe/H] $= -0.5$ is
reached (Romano et al. 2001). If they were to contribute significantly
to early $^7$Li
history, it would require the ad-hoc hypothesis that very strong mass loss (``super
winds'') occurs precisely at the moment when the surface of the star is lithium
rich (Ventura, D'Antona, \& Mazzitelli 2002). The low efficiency of AGB stars as
lithium factories is also supported observationally. North (1995) notices that
the barium-rich stars whose abundance patterns are explained by mass transfer
from an AGB companion are also lithium poor (A($^7$Li) $\leq$ 1.8 in most cases).
Finally the characteristic s-process pattern appears in halo stars once $\rm
[Fe/H] \geq -2.6$ (Truran et al. 2002, Simmerer et al. 2004, and references therein). This sugests
that AGB stars, the main sites for s-process nucleosynthesis, do not start to
affect the Galactic composition until a metallicity is achieved which
corresponds to objects on the Spite Plateau. If, despite these clues, we still
consider that the intermediate-mass stars could affect early lithium history we
must remember that in $\Lambda$-CDM cosmology the halo of the Galaxy is thought
to have formed from canibalization of smaller dwarf galaxies or an early
generation of globular clusters. For a 100 pc to 1 kpc sized dwarf galaxy, the
ISM should be rather well-mixed in a short amount of time, certainly less than
40 Myr (see \S \ref{sec5}). Therefore, no $^7$Li inhomogeneities are expected
because of the action of the first intermediate-mass stars. The third mechanism
that could be invoked for $^7$Li production is the supernova
$\nu$-process. It is associated with neutrino-wind induced reactions during
core-collapse supernovae (hereafter SNe). The Timmes et al. (1995)
chemical-evolution models suggest that the amount of $^7$Li produced by this
process is negligible in comparison to A($^7$Li) $\sim$ 2 at metallicities below
[Fe/H] $ = -1.0$. We hasten to add, however, that we have been made aware
of some recent computations of zero-metallicity core-collapse SNe that result in
higher lithium production than previously thought (A. Heger, private
communication). Hence, it may be premature to conclude that lithium production
cannot have had an early impact.

Bearing the above in mind, we neglected $^7$Li-production processes in our
discussion. Furthermore, we note that if these processes were to affect the early
Galactic history of $^7$Li one would expect significant variations in the
observed $^7$Li abundance at metallicities lower than the transition between
Population II (where A($^7$Li)$\rm \sim 2.1$) and the young Population I
low-mass stars (where A($^7$Li)$\rm \sim 3.2$). Indeed, even our value of $\rm
\bar{m}_{^7Li}$ is probably an upper limit. First, the stars in the range from
40 to 100 $\rm M_{\odot}$ will presumably lower $\rm \bar{m}_{^7Li}$ further, as
the trend of $\rm m_{^7Li}$ with mass in Figure \ref{fig4} already suggests.
Secondly, rotationally-induced mixing could bring the surface matter down into
regions of the star where lithium encounters proton capture.

The effects of rotation have just begun to be investigated in models of
Population III stars (Hirschi et al. 2006, Hirschi et al. 2006; Meynet,
Ekstr\"om, \& Maeder 2006). We draw attention to those models in which the
authors have followed the evolution of [Fe/H] $< -6$ massive (20-85 M$_\odot$),
rapidly rotating (300-800 km s$^{-1}$) stars over their relatively short
(several Myr) lifetimes. The most massive of these models ($\rm M_{star} > 40
M_\odot$) experience strong mass loss when they become red supergiants. They are
shown to have enriched surfaces in primary C, N and O (nitrogen is a species
whose production requires temperatures well above $2.5 \times 10^6$~K, hence it goes
along with lithium depletion). The net effect of internal rotationally-induced
mixing and mass loss is to inject prodigous amounts of C, N, and O into the
early ISM. This can raise the level of total metallicity to above the critical
value for low-mass star formation to occur. The stars formed from these material
will thus reflect the peculiar composition of these ejecta (see \S\ref{sec6}).
We will return to an examination of the possibility of rotationally-induced
mixing for explaining the unique composition patterns of HE~1327-2326 and
HE~0107-5240, but first we focus on the consequences of a low $\rm
\bar{m}_{^7Li}$ for Population III stars.

If we consider the predictions from the models for $^7$Li around the
main-sequence-turnoff for halo stars, we find that these objects have depleted
roughly 0.2 dex of their initial $^7$Li, mostly due to microscopic diffusion
and gravitational settling, as the non-standard turbulent-mixing near BCZ
can be considered as a secondary effect (Richard et al. 2005; Piau 2006). Since
the MS-turnoff Spite Plateau stars with metallicities [Fe/H] $\sim -3$ now
exhibit A($^7$Li)=2.1 to 2.2\footnote{There is still a debate about the exact
value of the Spite Plateau at [Fe/H] $\sim -3$. Some authors claim A($^7$Li)
$\sim$2.1 (Ryan et al. 1999), while others suggest A($^7$Li)$\sim$2.2 (Bonifacio
et al. 2006).The observations of the latter authors also suggest the
presence of turn-off halo stars that exhibit $\rm A(^7Li) < 2$, as is discussed in
\S\ref{sec2}.}, this means that their lithium abundances at the time of
their formation was nearly A($^7$Li) = 2.3 to 2.4. In turn, we can posit that
this lithium abundance is that which has resulted after mixing of Population III
processed ejecta with the BBN-composition ISM. If we take the BBN lithium
abundance to be A($^7$Li) = 2.6, we conclude that Population III stars are
responsible for a depletion on the order of $\rm \Delta ^7Li=$0.2 to 0.3 dex in
the very early epochs of the Galaxy. As we have shown that Population III
objects fully destroy their initial lithium, $\rm \Delta ^7Li$ directly
translates into the fraction of gas mass that was originally astrated by
Population III stars. Indeed, it suggests that between a third and a half of the
baryonic mass of the halo of the Galaxy was originally processed through
Population III objects. More generally, we can relate the fraction of mass $\rm
x_{PopIII}$ astrated by population III to $\rm \Delta ^7Li$ :

\begin{equation}
\rm x_{PopIII}=\frac{1-10^{-\Delta ^7Li}}{1-\bar{m}_{^7Li}}\sim 1-10^{-\Delta ^7Li}
\label{eq4}
\end{equation}

Equation \ref{eq4} is interesting, in the sense that it relates observed and
predicted quantities to the presently poorly-known efficiency of the Galaxy to
form its first generation of stars. The quantity $\rm \Delta ^7Li$ depends on
BBN, but is also closely related to our understanding of the internal dynamics
of low-mass Population II stars and their measured surface abundances, while
$\rm \bar{m}_{^7Li}$ is set by the structure of zero-metallicity stars and the
IMF in the early halo of the Galaxy. If we set $\rm \bar{m}_{^7Li}=6.7 \times 10^{-5}$
and $\rm \Delta ^7Li=0.3$ in equation \ref{eq4}, we obtain $\rm x_{PopIII}\sim
0.50$. If, at the other extreme, we set $\rm \bar{m}_{^7Li}=6.9 \times 10^{-3}$ and
$\rm \Delta ^7Li=0.2$ in equation \ref{eq4}, we obtain $\rm x_{PopIII}\sim
0.37$. The levels of astration we predict are quite important and should in turn
have consequences on (1) the early nucleosynthesis in the Galaxy, (2) the
injection of cosmic rays and subsequent light elements production in the ISM
when the (massive) Population III stars end their lives, and (3) the possible
reionization of the early interstellar/intergalactic medium. Although not the
main focus of the article, these three points are of cosmological significance,
and deserve some comments, as given below.

(1) Stellar nucleosynthesis. We predict that between a third and a half of the
matter constituting the present halo stars was astrated by Population III stars.
This large fraction seems worrying at first sight because it predicts a
large amount of early metal production. Some authors (Ricotti \& Ostriker 2004)
have argued that the enrichment in metals of the intergalactic medium (hereafter
IGM) by Population III stars constrains their masses or numbers. However, our
computations only imply that only up to half of the mass of the present
Population II stars was processed by Population III progenitors. This is much
smaller than the mass in the Galactic disk, and should not be confused with it
(see below). The SNe that take place at the end of the lives of massive
Population III stars may have occured at much lower masses than these stars
possessed originally, due to the effect of possibly large amounts of mass loss
from Population III massive stars during their lifetimes. Recent studies show
that they tend to reach their break-up velocities during the MS because of their
greater compactness, and their lower initial angular momentum loss through
magnetic winds than their Population I counterparts. Furthermore,
rotationally-induced mixing may, in the course of their evolution, bring metals
to the surface, which subsequently triggers stronger mass loss through stellar
winds (Hirschi et al. 2006; Meynet et al. 2006). Contrary to previous
assumptions, massive Population III stars would thereby lose a significant
amount of mass before exploding. The mass actually involved in the advanced
stages of quiescent or explosive burning would be much smaller than the ZAMS
mass. Thus, from the point of view of metal enrichment, the mass astrated by
Population III stars is smaller than half of the halo mass. It is finally
noteworthy that the wind composition in models of massive stars having $\rm
[Fe/H]=-6.6$ (Hirschi 2006) fits well the CNO abundance ratios observed in
HE~1327-2326 or HE~0107-5240, which are very different from the yields of
massive SNe, which exhibit much more oxygen than nitrogen or carbon (Hirschi et
al. 2006; Chieffi \& Limongi 2004). This suggests that the metals are less
efficiently ejected from the first SNe than they are from the following
generations.

Another possible issue raised by the hypothesized astration in Population III is
that, wherever $^7$Li is depleted, $^2$H is also depleted. We therefore predict
that between a third and a half of the primordial deuterium is already depleted
in the Population II stars. The primordial deuterium-to-hydrogen ratio expected
from WMAP constraints is $2.60\,10^{-5}$ (Coc et al. 2004), which is also in
good agreement with the measurements along the lines of sight to distant
quasars, $\rm (^2H/H)_p \sim 3 \times 10^{-5}$ (O'Meara et al. 2001; Burles 2002, and
references therein). On the other hand, the present deuterium-to-hydrogen ratio
in the local ISM lies around $1.5 \times 10^{-5}$ (Linsky 1998; Oliveira et al. 2003),
while the pre-solar ratio determined from the solar wind or the spectra of
gas-giant planets is between $2 \times 10^{-5}$ and $6 \times 10^{-5}$ (see Lemoine et al.
1999 for references). Knowing that deuterium is only destroyed by successive
generations of stars, both the $\rm ^2H/^1H$ ratios of the pre-solar
medium\footnote{The high $\rm ^2H/^1H$ ratio in the pre-solar medium deuterium
is problematic if the Galaxy evolves as a chemically isolated system.} and the
present local ISM should be smaller than $1.7 \times 10^{-5}$ ($1.5 \times 10^{-5}$) if one
third (respectively one half) of the matter of the Galaxy has been astrated by
Population III objects. However, the paradigm we suggest is that only
Population II objects result from a mix of Population III ejecta (and winds)
with genuinely BBN composition matter. This does not imply the same is true for
Population I objects. The infall of $^2$H-rich intergalactic matter of (near)
BBN composition onto the young Galactic disk may solve this apparent deuterium
discrepancy, provided it occurs after the formation of the Population II stars
and during the formation of the successive generation of Population I stars.
Such an infall scenario is indeed supported by the modeling of the deuterium
history in the Galaxy (Lubowich et al. 2000; Romano et al. 2006). Furthermore,
the stellar mass of the Galactic halo is $\rm \sim 2 \times 10^9 M_{\odot}$ (Bullock
\& Johnston 2005), i.e., only a few percent of the total Galactic mass. We
mention finally that the observed deuterium abundances in the present ISM
exhibit a large amount of scatter, which suggests a quite complex history for
this species. Accurate measurements remain a matter of debate (Lemoine et al.
1999; Romano et al. 2006).

(2) ISM nucleosynthesis. Since the seminal works of Reeves, Fowler, \& Hoyle
(1970) and Meneguzzi, Audouze, \& Reeves (1971), it has been known that
spallation and non-thermal fusion processes induced by cosmic-ray interactions
with the ISM can explain some of the observational patterns of the light
elements, namely Li, Be, and B over the history of the Galaxy. A discussion of
this topic is beyond the scope of the present work; for recent developments
about the LiBeB Galactic evolution through spallation we direct the reader to
Lemoine, Vangioni-Flam, \& Cass\'e (1998) and Fields \& Olive (1999). However,
we wish to make some brief remarks about the implications of our hypothesis of a
significant early astration for the $^6$Li isotope. 

Although $^6$Li has been detected for some time in halo turnoff stars (Cayrel et
al. 1999; Nissen et al. 1999), the sample of objects with claimed detections has
recently been increased significantly (Inoue et al. 2005; Asplund et al. 2006).
The observations of these authors suggest A($^6$Li)$\rm \sim 1$ is constant for
metallicities in the range $-2.7 \le {\rm [Fe/H]} \le -0.6$. Such a $^6$Li
plateau is puzzling because, contrary to $^7$Li, the amount of $^6$Li produced
by BBN should be negligible (Vangioni-Flam et al. 1999; see Jedamzik
2004 for a possible non-standard $^6$Li BBN scenario). However, the presence of
a $^6$Li plateau over such a broad range in metallicity suggests a pre-Galactic
origin. Making no assumptions on the mechanism(s) accelerating the
particles in the early ISM, Rollinde, Vangioni, \& Olive (2005) showed that a burst of
cosmological cosmic rays could explain the $^6$Li observed at very low
metallicity without overproducing $^7$Li. But what phenomenon produces the
energy required for fusion or spallation in the ISM? One mechanism that has been
proposed for accelerating the nuclei is related to the shocks associated with
structure formation (Suzuki \& Inoue 2002). This process was, however, recently
criticized by Prantzos (2006) because it produces too little energy, and would
lead to an increase of $^6$Li during a large fraction of the early Galactic
history, rather than to a plateau. Population III supernovae are alternate
candidates for $^6$Li production. They can inject the energy required for
the $\rm \alpha + \alpha$ fusion reactions thought to be the main channel of $^6$Li production in
the very metal-poor early ISM. In this respect, the large fraction of material
astrated by massive Population III stars we predict here is interesting,
because it supports a large number of early supernovae events. As previously
noted by Prantzos (2006), the extension of the $^6$Li plateau down to very low
metallicies places constraints on the nature of the early SNe, if they are
responsible for its production. In the case of the present-day composition
supernovae the ratio of kinetic energy (required to induce enough $^6$Li
production) to the iron injection in the ISM is such that Prantzos (2006)
establishes that it is not possible to produce A($^6$Li)$\rm \sim 1$ below
[Fe/H] = $-1.4$. However, the yields of Population III supernovae could differ
from those at present. As stressed by the same author (Prantzos 2005), the
electron mole fraction, the mass cut, and (above all) the rotationally-induced
effects have to be taken into account carefully before and during the explosion.
The uncertainties on these parameters leave Population III supernovae as
potential candidates for $^6$Li production at very low metallicity. Clearly,
this possibility should be explored more thoroughly in future works.

(3) Reionization. The reionization of the Universe can be tracked because it
affects the optical depth to the cosmic microwave background (Spergel et al.
2003). It is thought to have occured between a redshift of $\rm z=20$ and $\rm
z=6$ (Venkatesan 2006, and references therein); massive Population III stars
are strong candidates as the contributors of this ionizing radiation. If
reionization is related to the first stars, it is also necessarily connected to
Galactic chemical history. For instance, carbon and oxygen are observed in the
IGM and, provided the IMF and the chemical yields of Population III stars are
known, it is possible to constrain the reionization induced by the first stars
(Venkatesan \& Truran 2003). In our approach, $^7$Li also provides crucial
information about the efficiency of the first stars to convert baryonic matter
into IGM-ionizing photons, because it suggests that between one third and one
half of the baryonic matter of the halo ($\rm \sim 10^9 M_{\odot}$) was
processed in the first stars. Indeed, $^7$Li might be a more reliable probe
than other species because of its simpler evolution; it is systematically
destroyed in the astration process in a way that depends weakly on the IMF (see
above). On the contrary, the production of CNO and the other heavy elements
probably depends on the IMF, on the late stages of stellar evolution, and on the
complex and still to be fully investigated rotationally-induced mixing effects
(Hirshi et al. 2006a, 2006b).

\section{The Low Metallicity End of the Spite Plateau ?}\label{sec4}

As noted in \S \ref{sec2} above, several observed facts about the nature of the
Spite Plateau are of central importance to testing our Li-astration model.
First, the star-to-star scatter in measured A($^7$Li) for metal-poor stars on
the plateau is {\it extremely} small, on the order of 0.03-0.05 dex (Ryan et al.
1999; Asplund et al. 2006; Bonifacio et al. 2006), well within the expected
observational errors. Secondly, at the lowest metallicity end of the plateau,
$\rm [Fe/H] < -2.5$, there appears to be evidence for either a downturn in the
$^7$Li abundance relative to the more metal-rich stars ($-2.5 \le {\rm [Fe/H]}
\le -1.5$), or else an increase in the level of star-to-star scatter (Bonifacio
et al. 2006). This second point is dependent on observations of only a handful
of stars, and hence needs to be explored with much larger samples in the near
future. 

In our Li-astration model the downward trend (or increased scatter) of $^7$Li
abundances is interpreted as the imprint of the mixing between the Population
III ejecta and the ISM of BBN composition. We discuss the turbulent mixing of
the early Galaxy in the next section. Before the low-metallicity edge of the
plateau at $\rm [Fe/H] \sim -2.5$, not enough metals have been mixed into the
pure BBN-compostion ISM and the low-mass stars form more efficiently in the
unmixed ejecta because they contain the required amount of coolants, primarily C
and O. This stage extends from [Fe/H] = $-3.5$, which could be the lower metallicity
limit for low-mass star formation (Bromm \& Loeb 2003) up to [Fe/H] = $-2.5$. The
increase of scatter in r-process elements (Truran et al. 2002) favors this
hypothesis of still inhomogeneous ISM for metallicities lower than [FE/H] $\sim
-2.5$. However, we note that the scatter associated with r-process elements lingers
in the higher-metallicity regime at least up to $\rm [Fe/H] = -2$. For metallicities
in the range [Fe/H] $\sim -5$ to $\sim -3.5$, it has been very recently suggested by
Karlsson (2006) that the turbulence induced by the SNe explosions and the
ionizing effects of the massive first stars could lower the rate of star
formation. This would explain the apparent absence of stars on this metallicity
range. Interestingly, Karlsson interprets HE~1327-2326 the same way as we do, as a
hyper metal-poor star that formed predominantly out of the CNO-enriched ejecta
of a massive zero-metallicity star of the first generation. We stress that this
interpretation is in agrement with the yields of Population III models including
rotation mixing as mentioned by Hirschi et al. (2006) (see \S\ref{sec6})
. In this case HE~1327-2326 would be lithium poor for the same reason that it is
CNO rich -- because of astration of most of its mass by one (or a few) massive
Population III progenitor(s).

\section{Turbulent Mixing in the Early Galaxy}\label{sec5}

Observations of low-mass star formation in our Galaxy suggest that such 
stars form predominantly in giant molecular clouds on time scales
from $0.1$ to $\sim$ 1 Myr (Ward-Thompson, Motte, \& Andr\'e 1999). Observations
of these molecular clouds (Lis et al. 1996; Miesch, Scalo, \& Bally 1999) have
also shown that the gas within them is very turbulent. Similarly, observations
of starburst systems, such as M82 (Pedlar, Muxlow, \& Wills 2003), suggest that
star formation can take place in strongly shocked, turbulent environments with
very high rates of energy input from massive stars. In the case of early-epoch
galaxies there are also both theoretical and observational arguments to support
a similar connection between turbulence and star formation. We stress, however,
that the low-mass star formation in the present Galaxy cannot be transposed
without caution to the extremely metal-poor proto-Galactic ISM (Bromm \& Loeb
2003). From the modeling point of view, Elmegreen (1994) and Salvaterra,
Ferrara, \& Schneider (2004) suggested that converging shocks in a
proto-Galactic medium with high rates of supernova explosions might be one
possible mechanism for forming low-mass stars. From the observational point of
view the ULIRGs provide support for proto-Galactic ISMs that are strongly
influenced by massive stars and their ejecta. Therefore, in this section, we
take such a SN-driven proto-Galactic ISM as our working paradigm, and explore
metallicity evolution, lithium depletion, and low-mass star formation within
this context. Metal-poor stars, and in particular, hyper metal-poor stars,
provide essential clues about this early Galactic history.

It is beyond the scope of our present understanding to have a complete theory
for low- and high-mass star formation in our Galaxy. Nevertheless, some insights
may well be obtained. For example, if turbulence in the ISM is strong, and
mixing is extremely efficient over very short timescales, then it is reasonable
to expect that alpha-capture elements (which are made in all Type II SNe) should
be well-mixed into the proto-Galactic environment. Furthermore, if the
timescales required for complete mixing are comparable to the timescales for
low-mass star formation, then each generation of low-mass stars in the halo, as
cataloged by a specific value of [Fe/H], should also exhibit a small scatter in
the alpha-capture elements (see \S 6). On the other hand, it is also possible
that the earliest SNe were of the much rarer pair-instability type (Salvaterra
et al. 2004) and, as a result, the earliest ISMs were not strongly turbulent. In
that case we would expect hyper metal-poor low-mass stars to form only in the
high-metallicity ejecta left behind by such SNe. Then, the scatter expected in
metals in hyper metal-poor stars would be large, because they would have formed
in response to individual events, not all of which are identical.

If the primordial SNe were plentiful and evenly distributed we would expect the
proto-Galactic ISM to be strongly turbulent. In such a situation, the ejecta
from the posited Population III stars can be efficiently mixed with primordial
gas, resulting in a net dilution of the amount of $^7$Li in the natal gas clouds
that formed the next generation of stars (including low-mass stars). Theoretical
studies of turbulent mixing have been carried out by Bateman \& Larson (1993),
Roy \& Kunth (1995), and Oey (2003). The problem has also been studied
numerically by Klessen \& Lin (2003) and Balsara \& Kim (2005). The latter
authors were specifically interested in SN-driven turbulence, the very process
that interests us here. They found that for SN-driven turbulence that involves
a supernova rate that is eight times the present Galactic rate, the turbulent
diffusion coefficient is given by $\eta_{turb} = 5.7$ x $10^{26} \rm {cm}^2 /
{\rm sec}$. Since this is a diffusive process, the diffusion scales directly as
the square of the length of the system and inversely as the diffusion
coefficient. Therefore, we consider a 100 pc region, which could either be a
proto-globular cluster system or a part of a larger system, like M82, that is
undergoing an episode of vigorous star formation and SNe explosions. Taking $\rm
L_{100}$ to be a length measured in 100 pc and $\tau_{\rm diff,6}$ to be the
diffusion time, measured in millions of years, we obtain $\tau_{\rm diff,6} =
5.3 {\rm L_{100}^2\; Myr}$. This timescale is comparable to the total lifetime of
a massive star, with the consequence that each generation of stars that form do
so in an environment where the metals from the previous generation are
turbulently mixed in.

It is also important to realize that turbulence only mixes the metals at a
macroscopic level. An enhanced metallicity also produces an enhanced cooling
rate, thus playing an important role in regulating star formation. Since metals
play an important role in setting the cooling rate of gas, it is important to
mix the metals down to the molecular level. If such mixing is achieved for a
particular element, one would expect that element to show a very small scatter
with increasing [Fe/H]. The question of mixing the metals down to the molecular
level has been looked at theoretically by Oey (2003) and computationally by
Balsara \& Kim (2005). The latter authors found that such mixing occurs
unusually fast in turbulent environments. For example, in the case of the ISM
discussed above, they found that mixing of metals down to the molecular level
took place in a fraction of a million years. The implication is that, if 
turbulent eddies can efficiently mix the metals on the large eddy-bearing
scales, thus conveying supernova ejecta from one location to another, then the
ejecta will be mixed down to the molecular level in a time that is even shorter
than the bulk turbulent diffusion time. As a result, we anticipate that metals
that are produced in SNe of all mass ranges will be well mixed. The production
of scatter in the metals is a consequence of certain mass ranges of progenitor
stars producing certain species with above-average abundance, as well as
local pollution events (e.g., low-mass star formation in the near vicinity of a
given SNe, or mass-transfer events at later times).

The implication of the above discussion for $^7$Li is as follows. We see from
Figure \ref{fig2} that dwarfs with [Fe/H] $>-2.5$ exhibit an extremely small
scatter in A($^7$Li). A high rate of early turbulent mixing is indeed the most
natural way of explaining this low scatter. To obtain such a high rate of
turbulent mixing one requires a high rate of SNe, which is {\it a-posteriori}
justified if $\sim$30 to $\sim$ 50\% of the primordial matter of the halo was
processed by massive stars by the time [Fe/H] reached $\sim -2.5$ (\S 3).
Massive Population III stars not only explain the lowered mean value of
A($^7$Li) from the BBN value, because they deplete $^7$Li, but they also explain
the small scatter {\it above} [Fe/H] $\sim -2.5$ resulting from a very high SNe
rate and the subsequent vigorous ISM mixing at that epoch.
 
The perspective we have now developed would also enable us to understand the
trends in Figure \ref{fig2} for [Fe/H] $< -2.5$, where the scatter in A($^7$Li)
slightly increases with decreasing [Fe/H]. If the scatter keeps increasing with
decreasing [Fe/H], as more HE~1327-2326-like dwarfs are found, it would imply
that the hyper metal-poor dwarfs formed from mixtures of various
proportions between the metal-enhanced ejecta of Population III progenitors and
the BBN-composition ISM. It could also imply that these stars and the hyper
metal-poor dwarfs formed in response to individual supernova events, before a
sufficient number or rate of such supernova events could have produced
substantial amounts of turbulence in the proto-Galactic ISM. As mentioned
before, the scatter in r-process elements starting at $\rm [Fe/H]\sim -2$
and gradually increasing below $\sim$-2.5 indeed favors this possibility. If, on
the other hand, stars with [Fe/H] $< -2.5$ show a small scatter in $^7$Li, it
would imply that the ISM and SNe ejecta are efficiently mixed before these stars
form. Finally, if other hyper metal-poor stars such as HE~1327-2326 are found to
exhibit a complete absence of $^7$Li, it would imply that such stars cannot form
anywhere but within the ejecta of earlier massive stars.

\section{Additional Consequences of the Population III Processing model}\label{sec6}

In \S 3 we have drawn attention above to the recent models of Hirschi (2006),
Hirschi et al. (2006), and Meynet, Ekstr\"om, \& Maeder (2006), in which they
have followed the evolution of [Fe/H] $< -6$ massive (20-85 M$_\odot$), rapidly
rotating (300-800 km s$^{-1}$) stars. Such models experience
rotationally-induced mixing, which affects their nucleosynthesis and, in the most
massive ones ($\rm M_{\star} > 40 M_{\odot}$), triggers strong mass loss.
Eventually, prodigous amounts of C, N, and O are ejected into the ISM due to mass
loss or subsequent SNe explosions. This material will be barren of primordial
Li, but, as pointed by Hirschi et al. (2006), enriched in CNO to a level that could
allow low-mass star formation (Bromm \& Loeb 2003).

There are a number of other expectations for the stellar surface compositions of
the next-generation stars if rapidly-rotating zero-metallicity stars are
appealed to as the site of the Population III processing which we invoke for
lithium destruction. We discuss these in turn below.

\subsection{Unique Abundance Patterns Among Stars of the Lowest Metallicity}

The two known hyper metal-poor stars, HE~0107-5240 (Christlieb et al. 2002,
2004a), and HE~1327-2326 (Frebel et al. 2005; Aoki et al. 2006a), both with [Fe/H]
$< -5.0$, exhibit similar elemental abundance patterns, featuring enormous
overabundances (relative to Fe) of C, N, O, and (at least in the case of
HE~1327-2326), of the elements Na, Mg, and Al. Both of these stars, as well as
the two stars CS~22949-037 (Depagne et al. 2002) and CS~29498-043 (Aoki et al.
2002, 2004), with [Fe/H] $= -4.0$ (Cayrel et al. 2004) and [Fe/H] $= -3.5$ (Aoki
et al. 2004), respectively, are classified as Carbon-Enhanced Metal-Poor (CEMP)
stars, according to the suggested taxonomy of Beers \& Christlieb (2005). All
four are further sub-classified as CEMP-no stars, as none of them exhibit the
presence of Ba (note that only an upper limit, [Ba/Fe] $< 1.7$, exists for
HE~1327-2326). These broadly similar abundance patterns are thus far not seen
{\it in any other stars} with [Fe/H] $> -3.5$. Futhermore, Aoki et al. (2003)
and Aoki et al. (2006b) suggest that no more than 20\% of all CEMP
stars should be classified as CEMP-no stars, the remainder being of the class
that exhibit strong s-process (classified as CEMP-s) or r/s-process (classified
as CEMP-r/s) elemental abundance signatures.  

Predicted elemental abundance patterns that bear a strong similarity to the
stars discussed above are obtained in the ``wind only'' models described by
Meynet et al. (2006, their Figure 8; Hirschi 2006; Hirschi et al. 2006). In the view of our present
scenario, these four stars may well have recorded a nearly unaltered picture of
the material that was produced by their massive Population III zero-metallicity
progenitors.

\subsection{Production of Primary Nitrogen and the Observed Trends of [N/O]
and [C/O] at Low Metallicity}

Spite et al. (2005) have discussed the behavior of, in particular, nitrogen,
among the stars from the Cayrel et al. (2004) sample that are not expected to
have altered their surface values of CNO during their lifetimes (the ``unmixed
stars'' in the Spite et al. sample). They concluded that it was likely that some
source of primary nitrogen (that is, N that is formed directly from He, rather
than requiring conversion associated with CN processing) was necessary to
explain the observed high [N/O] (and [C/O]) ratios at low metallicity.
Chiappini, Matteucci, \& Ballero (2005) and, very recently, Chiappini et al.
(2006, based on the new yields computed by Meynet et al. 2006 and Hirschi et al.
2006 for rapidly rotating zero-metallicity massive stars) concluded that such
behaviors might be expected to follow if the Population III processing scenario
we consider in the context of $^7$Li astration applies in the early Galaxy. In
particular, the agreement of the new chemical evolution models from Chiappini et
al. (2006) with observations of the [N/O] and [C/O] trends at low metallicity is
quite striking. We recall here (see \S \ref{sec3}) that these ratios do not
correspond at all to those expected from typical Type II SNe ejecta.

\subsection{The Lack of Scatter in Alpha- and Iron-Peak Elements at Low Metallicity}

One of the clear outcomes of recent high-S/N, high-resolution investigations 
of very metal-poor stars (e.g., Carretta et al. 2002;  Cayrel et al. 2004; Arnone et
al. 2005) is the finding that an extremely small star-to-star scatter in the
observed alpha- and iron-peak elements exists for most stars with [Fe/H] $<
-2.0$. The implications of this result have been discussed by many authors. It
is difficult to imagine that, if Type II SNe with a wide range of initial masses
are responsible for the production of these elements, that there should be such
a small observed scatter. It seems, rather, that extremely efficient mixing, of
the sort we imagine (see \S 5) to explain the level of the $^7$Li Plateau, must have occured
in the ISM for at least the stars with [Fe/H] $> -2.5$. 

One might wonder about the observed (real) large star-to-star scatter in the
neutron-capture elements. In the view of our present investigation, these would
have to be accounted for by ``local'' pollution, either from binary companions,
e.g., AGB stars in the case of s-process-rich metal-poor stars, or the formation
of r-process-enhanced stars in the near vicinity of the objects responsible for
the production of the r-process elements.

\subsection{Carbon at Low Metallicity}

The Population III astration model, to which we appeal for lithium depletion,
is also a very attactive candidate for at least three carbon-related features
of very metal-poor stars with [Fe/H] $< -2.0$:

(i) One of the (at first) surprising results of the large modern surveys for
metal-poor stars is the apparent large increase in the fraction of
carbon-enhanced stars at low metallicity. Although there
remains some debate about the precise fraction of CEMP stars with [Fe/H] $<
-2.0$, recent results from analysis of high-resolution spectra from the HERES
survey (Christlieb et al. 2004b; Barklem et al. 2005) by Lucatello et al. (2006) make it clear
that at least 20\% of all stars at this low metallicity are enhanced in their
[C/Fe] ratios by a factor of ten or more above the solar value. Although the
sample size is presently small, Beers \& Christlieb (2005) point out that 5 of
the 12 stars known with [Fe/H] $< -3.5$, based on high-resolution
spectroscopic studies, are strongly carbon enhanced, roughly
40\%. Below [Fe/H] $= -5.0$, the fraction is 100\% (2 of 2 stars). It seems
required that some metallicity-dependent mechanism for the production of large
[C/Fe] ratios must be invoked. 

(ii) CEMP-s stars can be readily explained by appealing to the presence of a
binary companion that has donated s-process and carbon-rich material to the
surviving star that is presently observed, however there is no clear-cut model
for the production of carbon in the CEMP-no stars, where neutron-capture
elements are not found.

(iii) The isotopic ratio $^{12}$C/$^{13}$C in CEMP stars also provides
interesting constraints on the chemical history of the early halo. Although the
numbers of CEMP-no stars for which $^{12}$C/$^{13}$C has been measured thus far
is still small, it does appear that the observed isotopic ratio in these stars
is lower ( $^{12}$C/$^{13}$C $< 10$) than is typically found for the CEMP-s
stars (Beers et al. 2006; Sivarani et al. 2006).


The Population III processing model provides a natural explanation for these
features. In particular, no s-process-enriched material is expected to be
produced by the massive Population III stars, and it may well not be a
coincidence that the great majority of known CEMP-no stars have not yet
exhibited the presence of a binary companion from radial velocity monitoring
carried out to date. Furthermore the, ``wind only'' models of Hirschi et al.
(2006) and Meynet et al. (2006) produce very low $^{12}$C/$^{13}$C in the
material they eject prior to explosion; predicted ratios in the ejected material
are in the range $^{12}$C/$^{13}$C = 4--5. These change drastically (in their models) only post
explosion, rising by many orders of magnitude. The possible association of low
$^{12}$C/$^{13}$C with low Li is clearly of interest for future investigation,
in particular for the CEMP-no stars.

\section{Summary and Conclusions}\label{sec7}

We have addressed a new idea and considered its consequences regarding the
evolution of $^7$Li from Big Bang abundances to Population II halo star
abundances. This idea posits that the early lithium history lithium was
indeed significantly affected by astration of the matter {\it before} the
Population II stars were formed. The most compelling candidates for this task
are the massive Population III stars which, during their lifetimes, suffered
considerable mass loss of essentially Li-free and CNO-enhanced material, and
when they exploded, provided additional Li-free and metal-enriched matter to the
ISM. In this scenario the Population III ejecta and the ISM matter directly
inherited from the BBN are subsequently mixed by turbulence. Once enough metals
had been injected, the low-mass Population II stars we observe today began to
form.

\begin{figure*}[Ht]
\begin{center}
\includegraphics[angle=0,height = 6 cm, width=12 cm]{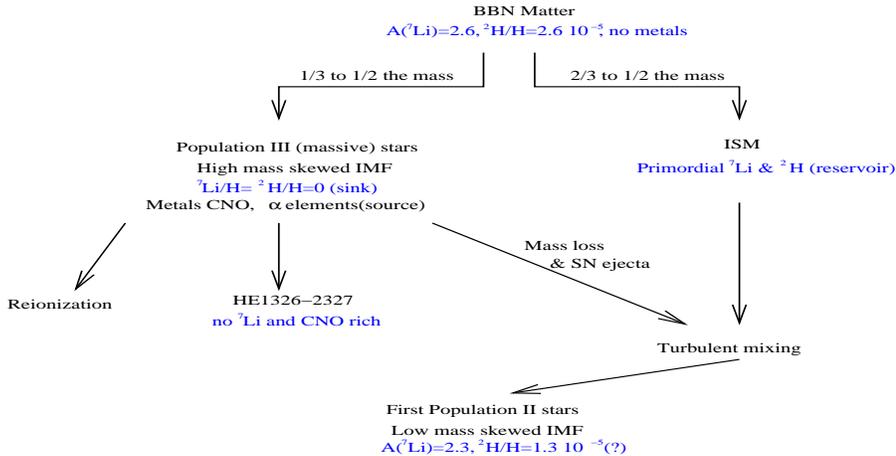}
\caption{Summary of the $^7$Li and heavier element 
chemical abundance evolution we suggest at the epoch of the Population III
stars. At that time the IMF is probably, contrary to now, skewed towards
high-mass stars.}
\label{fig5} 
\end{center}
\end{figure*}


Our starting point for this scheme was the apparent discrepancy between the high
primordial $^7$Li abundance inferred from BBN predictions and WMAP, A($^7$Li)
=2.6 (Coc et al. 2004), and the existing observations of MS-turnoff Population II
halo stars, A($^7$Li)=2.1 to 2.2 (see \S \ref{sec2}). Until now, this issue was
mostly addressed by considering the possible depletion of lithium {\it inside}
the Population II stars. A great number of works and various processes have been
invoked to deplete the surfaces of halo stars from their original lithium
abundances (e.g., Montalban
\& Schatzman 2000; Pinsonneault et al. 2002; Richard et al. 2005; Piau 2006). It
is noticeable, however, that explaining a $\rm \sim$0.4 dex discrepancy, together
with an extremely small scatter on the lithium plateau (around 0.02 dex
following Ryan et al. 1999 or Asplund et al. 2006), is a difficult task for any
of these processes. For instance, in a preceding paper, Piau (2006) reached the
conclusion that only $\rm 0.2$ dex of the initial $^7$Li was in fact depleted.
Similarly, Pinsonneault et al. (2002) estimated the depletion has to be in the
0.1 to 0.2 dex range, if one wants to be compatible with the (almost complete)
absence of scatter on the plateau. The recent detection of $^6$Li in a great
number of plateau stars (Asplund et al. 2006) further suggests that the $^7$Li
abundance observed there has mostly been affected by diffusion, but not by
nuclear destruction. Taking the recent Bonifacio et al. (2006) observations into
account raises the suspicion of the existence of a decrease in the $^7$Li
abundance when the metallicity drops below [Fe/H] = $-2.5$. This suggests
that the end of the Spite Plateau towards low metallicities has now been
reached, at roughly this metallicity. The absence of $^7$Li in HE~1327-2326, the
most metal-poor star presently known (Frebel et al. 2005), reinforces this
suspicion. 

We checked the predictions from stellar models of Population II turnoff stars
for the evolution of $^7$Li, assuming several prescriptions for non-standard
mixing and microscopic effects (i.e., microscopic diffusion and
gravitational settling; \S \ref{sec2}). Making the assumption that the present
observations of A($^7$Li) for turnoff stars are correct, we deduced that the
required depletion of lithium due to Population III processing lies roughly
between 0.2 and 0.3 dex (see Figure \ref{fig1} of the \S 1). This value
accounts for the amount of lithium destruction per solar mass astrated on the
ZAMS in Population III massive stars (\S \ref{sec3}). Let us now schematically
summarize our view of the temporal sequence of events marking the chemical
evolution of the early Galactic halo.

\subsection{Population III}

We interpret the present $^7$Li data in halo stars as follows.  The first
generation of stars forms, and roughly one third to one half of the baryonic mass of
the halo is processed through them (at least on the ZAMS). The species $^7$Li and $^2$H are
almost totally destroyed in these first stars. Because of the absence of metals
in the early ISM most (all?) of the first stars are massive -- the initial IMF
probably differs from the present IMF in the sense it is skewed towards larger
masses. Therefore, contrary to the present situation, most of the astrated mass
then resides in massive objects. These objects presumably lose a large fraction
of their mass due to rotational effects (\S \ref{sec3}), and evolve rapidly
into core-collapse supernovae. Globally, this process enriches the early ISM in
metals (CNO and $\rm \alpha$-elements; \S \ref{sec6}) while it depletes it of
light elements ($^7$Li and $^2$H; \S \ref{sec3}). Figure \ref{fig5} displays
this ``genealogy''. The evolution can perhaps be most easily thought of in terms of
``sinks'', ``sources'' and ``reservoirs''. Population III stars act as sinks for
$^7$Li and $^2$H, while they act as a sources for the heavy elements. The
ISM/IGM, on the other hand, acts as a reservoir for $^7$Li and $^2$H.

Since they are more metal rich, the winds of the massive Population III
stars and the subsequent SNe ejecta that are hardly mixed with the ISM are more
prone to form low-mass stars. In this respect, HE~1327-2326 could be one of the
objects primarily formed out of the winds and/or SNe ejecta of a Population
III star. No $^7$Li has been detected in its atmosphere, it has a very small
[Fe/H], but is CNO-rich, as expected from the nucleosynthesis of early
massive objects (Hirschi et al. 2006). In our framework, HE~1327-2326 differs
from the typical Population II objects in the sense that it was not formed from
a mix of early ISM matter and Population III ejecta. In some respects
HE~1327-2326 would be a transition object between Populations III and II. We
presented here the first isochrone for objects of the same composition as
HE~1327-2326 (\S \ref{sec2}).

We remark that the relation of Population III yields to the next generation of
stars might be a complex issue, and deserves further investigations (\S
\ref{sec5}). For instance, some segregation effects may have occured in the
ejecta, which could have affected the ratio between the different metals
(Venkatesan, Nath, \& Shull 2006). Furthermore, the shock waves of the first SNe
may have triggered low-mass star formation in a medium predominantly made of
metal-free matter (Salvaterra et al. 2004). Depending on the amount of mixing,
such low-mass Population II stars could simultaneously exhibit lithium lines and
extremely low ([Fe/H]$<$-3) metallicities. Investigating the A($^7$Li) scatter
in halo stars with [Fe/H]$<-2.5$ will provide important insights on this
question.

\subsection{Turbulent Mixing}

We have examined the question of turbulent mixing of metal-bearing supernova
ejecta in the proto-Galactic ISM. For high rates of SNe we find that the
turbulent diffusion of metals is very efficient. This is especially true in the
context of bottom-up, $\Lambda$-CDM cosmologies
suggested by WMAP. Efficient mixing of metals at and above [Fe/H] $\sim
-2.5$ is  a very economical strategy for explaining the small scatter in
A($^7$Li) at those metallicities. While the idea was not fully developed here,
the small scatter observed in alpha-capture elements by Cayrel et al. (2004)
also provides strong support for a scenario based on efficient turbulent mixing.

At values of [Fe/H] below $-2.5$ we expect that we should begin to probe the
effects of individual SNe, i.e., an insufficient sufficient number of SNe have
gone off in order to drive the turbulent mixing of the proto-Galactic ISM. In
those extremely early epochs we should expect to see the scatter in A($^7$Li)
increase with decreasing metallicity, provided $^7$Li is detected. Some of the
limited available data does exhibit this behavior, but better statistics are
required. Finally, we arrive at the interesting possibility that the ages of the
hyper metal-poor stars (if they could be readily measured) would provide an
alternative indication of the epoch of reionization. Our results also have
far-reaching consequences for cosmological simulations, because they imply that
turbulent processes may have been important at early epochs.

\subsection{Population II}

\begin{figure*}[Ht]
\begin{center}
\includegraphics[angle=0,height = 6 cm, width=12 cm]{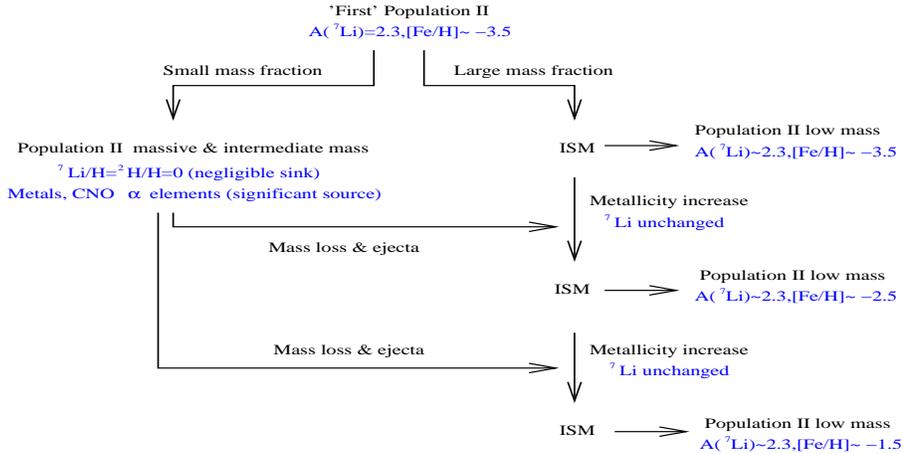}
\caption{Summary of the $^7$Li and metals 
chemical abundance evolution we suggest at the epoch of the Population II stars.
At that time the IMF is probably, as now, skewed towards low-mass stars.}
\label{fig6} 
\end{center}
\end{figure*}


As the turbulence in the Galactic medium mixes ever more Population III SNe
ejecta into the BBN-composition matter, the IMF becomes closer to what it is
today. Indeed, it is predicted that above a critical metallicity (around $\rm
Z=10^{-4}Z_{\odot}$, following Bromm \& Loeb 2003) low-mass stars begin to form.
Interestingly, the heavy-element mass fraction and [O/H] ratio we compute for
HE~1327-2326 are $\rm Z=3.1 \times 10^{-4}$ and $-1.8$, respectively, which is
comparable to the standard Spite Plateau composition where $\rm Z=3.0 \times
10^{-4}$, while [O/H] = $-1.7$ and [Fe/H] = $-2$. The change in the IMF is a
crucial point for the subsequent evolution of the baryonic matter in the halo --
by the time the low-mass stars gather most of the astrated matter, massive stars
no longer act as a significant sink for $^7$Li or $^2$H. On the main
sequence, the massive or intermediate-mass Population II stars deplete the light
elements very efficiently, as do their Population III counterparts. However, a
much smaller fraction of the astrated matter goes in massive or intermediate
mass stars once the average metallicity has increased to the Population II IMF
regime. Therefore, at this time, the global lithium destruction induced by
astration throughout the Galaxy should decrease.

Meanwhile, the massive stars keep on enriching the ISM with metals, because they
evolve rapidly. We should therefore expect a nearly constant value of the $^7$Li
abundance, despite the increase of metallicity or even a small increase, because
of cosmic-ray $^7$Li production due to spallation or moderate SNe $^7$Li production.
In the present framework, the slightly more lithium-poor stars observed by
Bonifacio et al. (2006) and Asplund et al. (2006) around [Fe/H] = $-2.5$ mark
the low-metallicity end of the lithium plateau. This may also be seen as the
last stages of turbulent mixing ,before the Population III SNe ejecta and the
primordial ISM are mixed to the point that the early halo becomes nearly
chemically homogeneous. Figure \ref{fig6} illustrates the evolution we propose,
once the metallicity is high enough to form Population II low-mass stars.

\subsection{Final Remarks}

Our scenario is still rather speculative, and has to be checked by further
observational and theoretical efforts. However, we stress that a
significant astration of the halo matter by Population III stars could
simultaneously solve three present issues related to $^7$Li: (1) the absence of
this isotope in the hyper metal-poor star HE~1327-2326, (2) the apparent edge of the
Spite Plateau around [Fe/H] = $-2.5$, and (3) the discrepancy between current BBN
predictions of primodial lithium and the observations of lithium abundances on
the Spite Plateau. A 30\% to 50 \% astration fraction of halo matter in
Population III stars is certainly a high value (see \S \ref{sec3}). Should this
astration fraction be strongly reduced, the questions raised by HE~1327-2326 and
the edge of the Spite Plateau could still be related to Population III stars.
One major ingredient for checking the present scenario is the chemical
yields of the first stars. The rotationally-induced mixing effects have been
demonstrated to explain many stellar physics features during the quiescent phase
of nucleosynthesis at non-zero metallicity. These effects, as well as the explosive
nucleosynthesis of the first SNe, need to be investigated further at zero
metallicity. These new models should, in particular, provide an estimate of the
global metal yields of the first stars in fair agreement with the observations
of hyper metal-poor stars. They should also be compared with the various
distinctive chemical features associated with carbon and nitrogen trends among
the oldest stars of the halo. In parallel, it is important to model further the
lithium evolution in hyper metal-poor stars, and to investigate the question of
microscopic diffusion and gravitational settling in more detail in
this environment. Additional $^7$Li observations of extremely, ultra, and
hyper metal-poor stars are clearly needed as well, for several reasons. First,
in order to confirm or refute the possible change of slope in the A($^7$Li)-[Fe/H] relation around
[Fe/H] = $-2.5$, secondly, to inspect the $^7$Li abundances for dwarf stars in
the same metallicity range as HE~1327-2326. The question of whether the stars
in the hyper metal-poor metallicity range are predominantly lithium poor or
lithium rich would, respectively, reinforce or weaken the scenario we discuss in
this paper.

Another major ingredient required for exploring the scenario we suggest is
better understanding of the efficiency of turbulent mixing, and its ability to
homogenize the early ISM during the time when tee first genrations of stars were
forming. This is clearly a very complex issue given the interactions between the
stellar winds and ejecta, the stellar ionization effects and the interstellar
magnetic fields. This point is also connected to the scenario for the formation
of the Galaxy, which requires further modeling efforts. The elemental abundance
patterns of the first stars can provide crucial observational clues to all of
these issues. For the halo stars that exhibit [Fe/H] $> -2.5$, it is tempting to
interpret the low scatter in $^7$Li, the $\alpha$-elements, and the iron-peak
elements to an imprint arising from early, efficient, mixing of the ISM.
Below this metallicity, the scatter in the $\alpha$-elements and the iron-peak
elements remains small (Cayrel et al. 2004). The scatter in r-process
elements seems to increase for metallicities below $\rm [Fe/H] = -2$ (Truran et
al. 2002). Understanding these apparently contradictory behaviors would provide
crucial information about the competing efficiencies of the mixing and the
stellar formation processes at the earliest epochs of the halo of the Galaxy.

If our general picture is correct, deuterium is depleted in Population II stars
by at least a factor 1.5--2 below the BBN deuterium abundance. This by no
means implies that half the matter of the entire Milky Way disk has been
astrated in Population III stars, nor that pre-solar or local ISM $\rm
^2H/^1H$ ratio should be less than a third or a half of its primordial value.
The stellar mass of the halo is roughly two orders of magnitude {\it below} the
stellar mass of the Galaxy. Moreover, the present models and observations of
deuterium in the Galactic disk suggest that replenishment of this isotope has taken
place, ``likely from infall of gas of cosmological composition'' (Romano et al.
2006). These authors predict that the deuterium fraction is constant, and near
the primordial fraction, during Galactic evolution from [O/H] = $-3$ to
[O/H] = $-0.5$. Since deuterium is more easily destroyed by astration than
lithium, the results of Romano et al. suggests that no $^6$Li or $^7$Li depletion
occurs in the ISM for the metallicity range of the Spite Plateau stars.

\begin{acknowledgements}

L.P. and J.W.T. acknowledge support from the National Science Fundation under
Grant PHY 02-16783 for the Physics Frontier Center Joint Institute for Nuclear
Astrophysics (JINA). J.W.T. also acknowledges support from the US-DOE, Office of
Nuclear Physics, under contract No. W-31-109-ENG-38. T.C.B. and S.T. acknowledge
partial funding for this work from grants AST 04-06784 and PHY 02-16783: Physics
Frontiers Center/Joint Institute for Nuclear Astrophysics (JINA), both awarded
by the U.S. National Science Foundation.

\end{acknowledgements}

\end{document}